\documentclass[prl,twocolumn]{revtex4-1}
\usepackage{amsmath}
\usepackage{amssymb}
\usepackage{array}
\usepackage{mathrsfs}
\usepackage{tabularx}
\usepackage{graphicx}
\usepackage{dcolumn}
\usepackage{bm}
\usepackage{graphicx}
\usepackage{color}
\usepackage{booktabs}
\usepackage{multirow}
\usepackage{diagbox}

\begin{document}
\title{Experimental measurement-device-independent quantum key distribution with the double-scanning method}
\author{Yi-Peng Chen$^{1,2,3,\dagger}$}
\author{Jing-Yang Liu$^{1,2,3,\dagger}$}
\author{Ming-Shuo Sun$^{1,2,3}$}
\author{Xing-Yu Zhou$^{1,2,3,\S}$}
\author{Chun-Hui Zhang$^{1,2,3}$}
\author{Jian Li$^{1,2,3}$}
\author{Qin Wang$^{1,2,3}$}

\email{xyz@njupt.edu.cn}
\email{qinw@njupt.edu.cn}

\affiliation{$^{1}$Institute of quantum information and technology, Nanjing University of Posts and Telecommunications, Nanjing 210003, China.}
\affiliation{$^{2}$"Broadband Wireless Communication and Sensor Network Technology" Key Lab of Ministry of Education, NUPT, Nanjing 210003, China.}
\affiliation{$^{3}$"Telecommunication and Networks" National Engineering Research Center, NUPT, Nanjing 210003, China.}

\begin{abstract}
\noindent The measurement-device-independent quantum key distribution (MDI-QKD) can be immune to all detector side-channel attacks. Moreover, it can be easily implemented combining with the matured decoy-state methods under current technology. It thus seems a very promising candidate in practical implementation of quantum communications. However, it suffers from severe finite-data-size effect in most existing MDI-QKD protocols, resulting in relatively low key rates. Recently, \textit{Jiang et al.} [Phys. Rev. A 103, 012402 (2021)] proposed a double-scanning method to drastically increase the key rate of MDI-QKD. Based on \textit{Jiang et al.'s} theoretical work, here we for the first time implement the double-scanning method into MDI-QKD and carry out corresponding experimental demonstration. With a moderate number of pulses of $ 10^{10} $, we can achieve 150 km secure transmission distance which is impossible with all former methods. Therefore, our present work paves the way towards practical implementation of MDI-QKD.

\end{abstract}

\maketitle
\section{Introduction}
Quantum Key Distribution (QKD), based on the laws of quantum physics, can offer unconditional secure communication between two legitimate parties (Alice and Bob) \cite{Artur,Lo,May}, even if there exists a malicious eavesdropper (Eve). The first QKD protocol is proposed by Bennett and Brassard in 1984 named BB84 \cite{Bennett}. Hereafter, it has attracted extensive attention and developed rapidly both in theory and experiment \cite{I6,I7,I8}. The security proof of BB84 protocol with nonideal photon sources was given by Gottesman et al. \cite{Single1,Single2}, and the decoy state method \cite{Wang,LO. HK} was invented to counter attack the photon number splitting (PNS) attacks \cite{PNS}, corresponding experimental demonstrations were exhibited \cite{Tokyo,BB84 421,Onchip}. Moreover, to solve those attacks directed on the detecting side, the measurement-device-independent quantum key distribution (MDI-QKD) protocol was put forward \cite{MDI1,MDI2}. Combined with the decoy-state method, MDI-QKD has the ability of avoiding attacks aiming at detectors and multi-photon components in the light source. Thereafter, a larger number of theoretical and experimental works have been carried out to improve practical performances of MDI-QKD \cite{MDIImprove3,MDIImprove4,MDIImprove6,MDIImprove7,MDIImprove8,MDIImprove9}. 

Up to date, the secure transmission distance of MDI-QKD has been extended up to more than four hundred kilometres \cite{MDIImprove6} which shows great advantages in long-distance communication. However, its key rate is still seriously affected by the finite-size effect, and the block size used for considering finite-size effects is usually larger than $ 10^{12} $ \cite{MDIImprove6,MDIImprove9}. Fortunately, the 4-intensity MDI-QKD protocol \cite{MDIImprove4} and the double-scanning method \cite{DoubleScan} have been invented and dramatically improve the performance of the MDI-QKD. Based on the above theoretical work, we carry out a proof-of-principle experimental demonstration, and mainly focus on the situation with a small data size, i.e., $ 10^{10} $ pulses are emitted and 150 km transmission distance is successfully achieved. 

In what follows, we first briefly review the theory on the MDI-QKD protocol with double-scanning method in Sec. \uppercase\expandafter{\romannumeral2}. We then describe our decoy-state MDI-QKD experimental setup in Sec. \uppercase\expandafter{\romannumeral3}. Experimental results are shown in Sec. \uppercase\expandafter{\romannumeral4}. Finally, the conclusion and outlook are given out in Sec. \uppercase\expandafter{\romannumeral5}.

\section{Theory}

In the following, we will briefly review the implementation flow of the 4-intensity MDI-QKD protocol with the double-scanning method. 

$Step$ (i). Alice (Bob) randomly modulates the light source $l$ $(r)$ into four different intensities, including the signal state $\mu$ , the decoy states $\nu$, $\omega$ and the vacuum state $o$, i.e., $l$ $(r) \in \{\mu, \nu, \omega, o\}$. The probabilities of choosing different intensities for Alice (Bob), are denoted as $P_l$ ($P_r$). The photon-number distribution of the phase-randomized light sources follows the Poisson distribution, $p_{n}^{\lambda} = \frac{\lambda^{n}}{n!}e^{-\lambda}$, where $\lambda$ is the average intensity, and $n \in \{0,1,2...\}$. In our work, encoding bases are decided by the different intensities, and the signal state pulses are only prepared in the Z basis while the decoy-state pulses are modulated in the X basis.

$Step$ (ii). Charlie performs Bell-state measurements (BSMs) on the incident pulse pairs sent by Alice and Bob. For simplicity, we only consider the effective event when the pulse pairs are projected onto state ${|\psi^- \rangle}=({|01 \rangle}-{|10 \rangle})/\sqrt{2}$, where Alice and Bob can easily make their bit strings identical through fliping any one’s bit. After sufficient effective-event counts are recorded, Charlie publicly declares the BSM results. 

$Step$ (iii). Alice and Bob announce their basis choices to implement basis reconciliation. Corresponding gains $Q_{lr}^{XX}$, $Q_{lr}^{ZZ}$ and the quantum bit error rates (QBER) $E_{lr}^{XX}$, $E_{lr}^{ZZ}$ can be obtained directly. With these experimental observations, we carry out parameter estimations. Then, error correction and privacy amplification are processed to obtain the final secure key strings.

In the process of parameter estimation, we need to calculate the yield of single-photon-pair contributions $(Y_{11}^{ZZ})$ and the phase-flip error of single-photon-pair $(e_{11}^{ph})$ in signal states. According to Ref. \cite{MDIImprove1} we have the lower bound of single-photon-pair yield $(Y_{11,L}^{XX})$ and upper bound of the bit-flip error rate $({e_{11,U}^{XX}})$ in decoy states:
\begin{equation}
\begin{split}
{Y_{11,L}^{XX}} \geq  &\dfrac{1}{{p_1^\omega}{p_1^\nu}({p_1^\omega}{p_2^\nu}-{p_1^\nu}{p_2^\omega})}[({p_1^\nu}{p_2^\nu}{\langle Q_{\omega\omega}^{XX}\rangle}\\
&+{p_1^\omega}{p_2^\omega}{p_0^\nu}{\langle Q_{o\nu}^{XX}\rangle}
+{p_1^\omega}{p_2^\omega}{p_0^\nu}{\langle Q_{\nu{o}}^{XX}\rangle})\\
&-({p_1^\omega}{p_2^\omega}{\langle Q_{\nu\nu}^{XX}\rangle}+{p_1^\omega}{p_2^\omega}{p_0^\nu}{p_0^\nu}{\langle Q_{o{o}}^{XX}\rangle})\\
&-{p_1^\nu}{p_2^\nu}({p_0^\omega}{\langle Q_{o{\omega}}^{XX}\rangle}+{p_0^\omega}{\langle Q_{{\omega}o}^{XX}\rangle}\\
&-{p_0^\omega}{p_0^\omega}{\langle Q_{oo}\rangle})],
\end{split}
\end{equation}

\begin{equation}
\begin{split}
{e_{11,U}^{XX}} \leq 
&\dfrac{1}{{p_1^\omega}{p_1^\omega}{Y_{11}^{XX}}}[{\langle Q_{\omega\omega}^{XX}E_{\omega\omega}^{XX} \rangle}-({p_0^\omega}{\langle Q_{o\omega}^{XX}E_{o\omega}^{XX} \rangle}\\
&+{p_0^\omega}{\langle Q_{\omega{o}}^{XX}E_{\omega{o}}^{XX} \rangle}
-{p_0^\omega}{p_0^\omega}{\langle Q_{oo}E_{oo}\rangle})],
\end{split}
\end{equation} where ${\langle * \rangle} $ represents the expected value of the experimental observation. Refering to the double-scanning method \cite{DoubleScan}, we extract those common parts that exists in ${Y_{11,L}^{XX}}$ and ${e_{11,U}^{XX}}$, i.e., the error counts rate $\mathcal{M}={\langle Q_{\omega\omega}^{XX}E_{\omega\omega}^{XX} \rangle}$ and the vacuum related counts rate $\mathcal{H}={p_0^\omega}{\langle Q_{o\omega}^{XX} \rangle}+{p_0^\omega}{\langle Q_{\omega{o}}^{XX} \rangle}-{p_0^\omega}{p_0^\omega}{\langle Q_{oo}\rangle}$. Thus, we can reformulate Eqs. (1) and (2) as:
\begin{equation}
\begin{split}
{Y_{11,L}^{XX}} \geq &\dfrac{1}{{p_1^\omega}{p_1^\nu}({p_1^\omega}{p_2^\nu}-{p_1^\nu}{p_2^\omega})}[({p_1^\nu}{p_2^\nu} \bar {\rm {\mathcal{M}}}+{p_1^\omega}{p_2^\omega}{p_0^\nu}{\langle Q_{o\nu}^{XX}\rangle}\\
&+{p_1^\omega}{p_2^\omega}{p_0^\nu}{\langle Q_{\nu{o}}^{XX}\rangle})-({p_1^\omega}{p_2^\omega}{\langle Q_{\nu\nu}^{XX}\rangle}\\
&+{p_1^\omega}{p_2^\omega}{p_0^\nu}{p_0^\nu}{\langle Q_{o{o}}^{XX}\rangle})
+{p_1^\nu}{p_2^\nu}(\mathcal{M}-\mathcal{H})],
\end{split}
\end{equation}

\begin{equation}
\begin{split}
{e_{11,U}^{XX}} \leq \dfrac{1}{{p_1^\omega}{p_1^\omega}{Y_{11}^{XX}}}(\mathcal{M}-\mathcal{H}/2),
\end{split}
\end{equation} where $\bar {\rm {\mathcal{M}}}\ = {\langle Q_{\omega\omega}^{xx}(1-E_{\omega\omega}^{xx})\rangle}$,
and the expected values in the above formulas can be calculated with the observed values through Chernoff bound method \cite{chernoff}:

\begin{equation}
\begin{split}
&{\langle {\Gamma}_{lr} \rangle} \geq {F^L({\Gamma}_{lr})} =: {\Gamma}_{lr} - f((\frac{\epsilon}{2})^{3/2})\sqrt{\dfrac{{\Gamma}_{lr}}{N_{lr}}},\\
&{\langle {\Gamma}_{lr} \rangle} \leq {F^U({\Gamma}_{lr})} =: {\Gamma}_{lr} + f(\dfrac{(\epsilon/2)^4}{16})\sqrt{\dfrac{{\Gamma}_{lr}}{N_{lr}}},
\end{split}
\end{equation} where ${\Gamma}_{lr}$ denotes the experimental observation and $f(x) = \sqrt{2ln(1/x)}$. $F^L(*)$ and $F^U(*)$ represent the lower and upper of Chernoff bound, respectively. Here $\epsilon$ is the failure probability \cite{MDISecure}.

Moreover, we combine the technique of jonit constrains \cite{DoubleScan} and Eq. (5) to construct linear programming of the $Y_{11,L}^{XX}$ related to $\bar {\rm {\mathcal{M}}}$, ${\langle Q_{o\nu}^{XX}\rangle}$, ${\langle Q_{\nu{o}}^{XX}\rangle}$, ${\langle Q_{\nu\nu}^{XX}\rangle}$ and ${\langle Q_{o{o}}^{XX}\rangle}$. By solving the problem of linear programming \cite{DoubleScan}, we can get the estimation values of $Y_{11,L}^{XX}$ and $e_{11,U}^{XX}$. 

Furthermore, according to the proof in Ref. \cite{MDIImprove4}, the yield ${Y_{11,L}^{ZZ}}$ and  the phase-flip error $e_{11,U}^{ph}$ in signal states satisfy: 
\begin{equation}
\begin{split}
{Y_{11,L}^{ZZ}} = {{Y_{11,L}^{XX}}},\qquad \qquad 
\end{split}   
\begin{split}
e_{11,U}^{ph} = {e_{11,U}^{XX}}.
\end{split}
\end{equation}

With the above estimated parameters, the secure key generation rate can be calculated:
\begin{equation}
\begin{split}
R(\mathcal{H},\mathcal{M}) = {{p_{\mu}}^2{\left\{{(p_1^{\mu})^2}{Y_{11,L}^{ZZ}}{[1-h(e_{11,U}^{ph})]-{fQ_{{\mu}{\mu}}^{ZZ}h(E_{{\mu}{\mu}}^{ZZ})}}\right\}}}
\label{ep}
\end{split} 
\end{equation} where $Q_{{\mu}{\mu}}^{ZZ}$ is the overall gain and $E_{{\mu}{\mu}}^{ZZ}$ is the QBER in signal states, and their values can be observed in experiment; $h(x) = -x{\log_2(x)}-(1-x){\log_2(1-x)}$ is the binary Shannon information function. 

Moreover, the lower and upper bounds of $\mathcal{H}$ and $\mathcal{M}$, i.e., $[{\mathcal{H}_L},{\mathcal{H}_U}]$ and $[{\mathcal{M}_L},{\mathcal{M}_U}]$, can also be acquired with the process of joining constraints described above. Then, we simultaneously scan $\mathcal{H}$ in $[{\mathcal{H}_L},{\mathcal{H}_U}]$ and $\mathcal{M}$ in $[{\mathcal{M}_L},{\mathcal{M}_U}]$ to implement the joint study, and get the final key rate:
\begin{align}
R_L = \min_{{\mathcal{H} \in [{\mathcal{H}_L},{\mathcal{H}_U}]}\atop{\mathcal{M} \in [{\mathcal{M}_L},{\mathcal{M}_U}]}}R(\mathcal{H},\mathcal{M}).
\label{ep}
\end{align}

\section{Experiment}

The schematic of our experimental setup is shown in Fig. 1. In this work, we employ Faraday-Michelson interferometers (FMIs) to implement a time-bin phase encoding scheme. Alice and Bob each have a narrow linewidth continuous-wave laser (Clarity NLL-1550-LP), whose wavelength is locked to the P14 line of C13 acetylene at 1550.51 nm. The frequency-locked lasers ensure that the two-photon interference at Charlie's side is consistent in frequency. Four intensity modulators (IMs) are deployed on each side, and we will singly introduce the function of the following IMs from the source to the detectors. The first two IMs are applied to chop continuous light into a pulse train with a repetition rate of 50 MHz and further modulate them into the decoy states. FMI, as the core device of the coding scheme, is composed of a 50/50 beam splitter (BS), a phase modulator (PM) and two Faraday mirrors (FMs). Moreover, the circulator can effectively filter light pulses reflected by the FMs. Each pulse entering the FMI is split into front and rear time bins. The arm length difference between the long and short arms of the FMI distinguishes time bins with a temporal difference of 10.3 ns. Phase encoding in the X basis can be performed by changing the extra phase voltages of PM at the long arm. The following third and fourth IMs are adopted for basis choice and time-bin encoding. For the Z basis, the two IMs can remove the front or rear time bin, representing $|0\rangle$ or $|1\rangle$ bit respectively. The cascaded IMs can further improve the extinction ratio, giving a reduced optical error rate (~0.1\%) in the Z basis. Moreover, considering the PM inevitably causes insertion loss, the latter two IMs should also designed to balance the intensity difference between the front and rear bin pulses. To obtaining the optimal operating voltages of PMs and IMs, periodically scanning the voltages and analyzing corresponding count rates are necessary for keeping free running experimental systems.
\begin{figure*}
	\centering
	\includegraphics[width=6.5in]{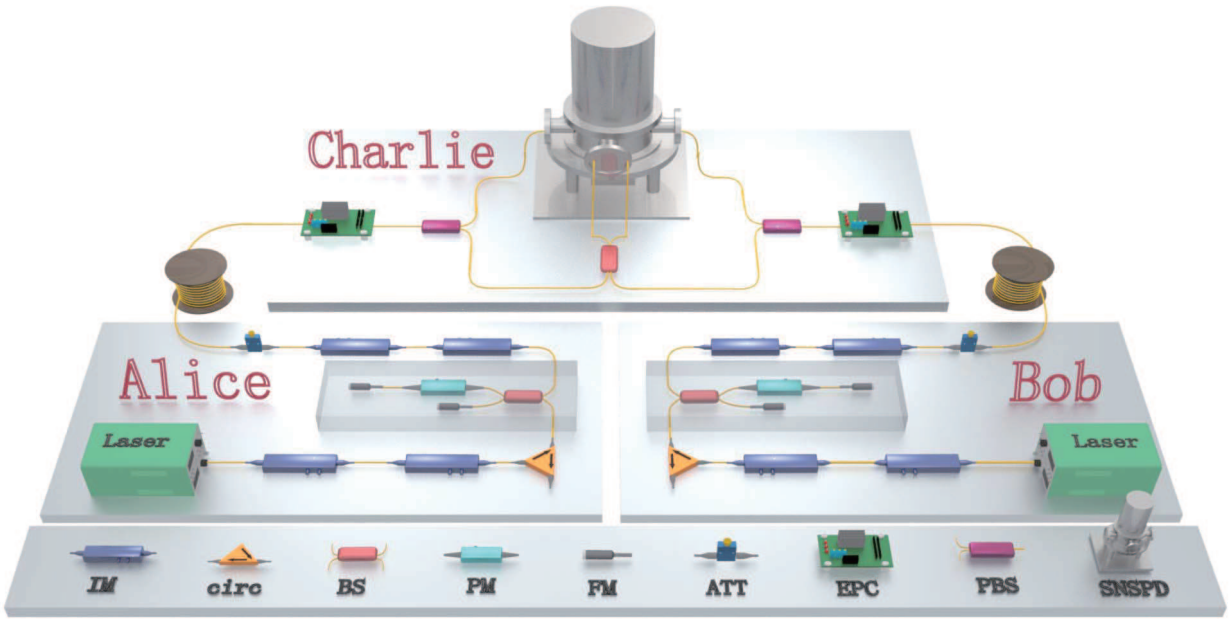}
	\caption{Schematic of the experimental setup. Laser: continuous-wave laser; IM: intensity modulator; Circ: circulator; BS: beam splitter; PM: phase modulator; FM: Faraday mirror; ATT: attenuator; EPC: electronic polarization controller; SNSPD: superconducting nanowire single-photon detector. }
	\label{experiment}
\end{figure*}
The light pulses are attenuated to single-photon level before sent to Charlie through the commercial standard single-mode fiber with a transmission coefficient of 0.18 dB/km. Then, the pulse pairs from Alice and Bob interfere at Charlie's BS, further sent into two commercial superconducting nanowire single-photon detectors (SNSPDs) which are connected to the time-to-digital converter (TDC) for data processing. To be specify, the quality of two-photon interference is determined by the photon frequency, the arrival time and the photon polarization. Here the temporal indistinguishability has been kept by inserting an electric optical delay line in Alice's side (not shown in Fig. 1). To maintain the polarization stabilization, two electronic polarization controllers (EPC) plus two polarization BSs (PBSs) are inserted before the BS to implement polarization automatic compensation. Finally, the interference visibility in our experiment is better than 48\%. 

The total efficiency of the experimental system is 60\% including the losses of EPC, PBS, BS, and SNSPDS. Moreover, the SNSPDs used in this work run at 2.2 K and provide 85\% detection efficiencies at dark count rates of 12 Hz.

\section{Results and discussion}
\begin{figure}[h]
	\begin{center}
		\includegraphics[width=3.5in]{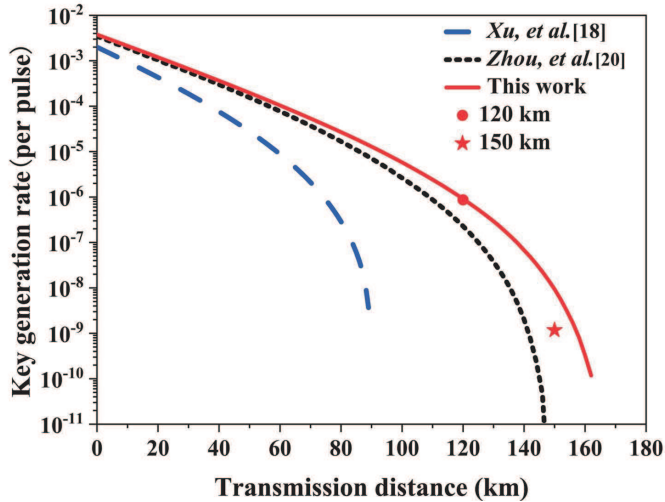}
	\end{center}
	\caption{The theoretical and experimental secret key rates versus the transmission distance. The red solid line represents our theoretical key rates while the dot point and star point represent experimental results with 120 km and 150 km fibers, respectively; The black dotted line indicates theoretical predictions of Ref. \cite{MDIImprove4} and the blue dash line represents Ref. \cite{MDIImprove1}.}
	\label{fig-2}
\end{figure}
The system parameters implemented in our experiment are listed in Table 1. The total number of pulses used in the experiment is $10^{10}$, and the failure probability $\epsilon$ is reasonably set as $10^{-10}$. The calculated optimal experimental parameters and the relevant data are displayed in Table 2. We conduct the experiment over 120 km and 150 km fibers, and achieve corresponding key rates as 43.54 bps and 0.06 bps respectively, which agrees well with theoretical predictions. Moreover, in order to illustrate the advantages of the new scheme, we also plot out the variation of the key rate with transmission distance by using either 3-intensity \cite{MDIImprove1} or 4-intensity \cite{MDIImprove4} decoy-state method, see Fig. 2. Obviously, the present double-scanning method shows overwhelming advantage compared with the other two schemes.
\begin{table}[ht]
	\renewcommand\arraystretch{1.3}
	\caption{List of experimental parameters. Here $\alpha$ is the fiber loss coefficient (dB/km); $\eta_{d}$ is the total efficiency of the Charlie; $Y_{0}$ is the dark count rate of the Charlie's detectors; $e_{d}^{Z}$ and $ e_{d}^{X}$ are the misalignment-error probability on the Z and X bases, respectively; $f$ is the inefficiency of error correction; $\epsilon$ is the failure probability.}
	\setlength{\tabcolsep}{2.0mm}
	\begin{tabular}{ccccccc} \hline\hline
		$\alpha$ &$\eta_{d}$&$Y_{0}$&$e_{d}^{Z}$ &$ e_{d}^{X}$&$f$&$\epsilon$ \\ \hline
		0.18 dB/km&$60\%$&$4\times10^{-8}$&0.1\%& 1\%&1.16& $10^{-10}$  \\
		\hline  \hline
		\label{Parameter}
	\end{tabular}
	\label{table-1}
\end{table}

\begin{table}[h]
	\renewcommand\arraystretch{1.5}
	\caption{The optimal parameters of corresponding distance. $\mu$ is the intensity of signal state. $\nu$, $\omega$ are the intensities for decoy states. $P_{\mu}$, $P_{\nu}$, $P_{\omega}$ are the probabilities to choose different intensities.}
	\setlength{\tabcolsep}{2.25mm}
	\begin{tabular}{ccccccc}  \hline  \hline
		&  $\mu$ & $\nu$ & $\omega$ &$P_{\mu}$ & $P_{\nu}$ & $P_{\omega}$  \\ \hline
		120&0.5866	&0.3323&0.0767&	 0.4151&0.1337&0.4305\\
		150 & 0.3851	&0.3707&	0.0763& 0.1763&0.1898&0.6124\\
		\hline  \hline
	\end{tabular}
	\label{table-2}
\end{table}

\section{Summaries and outlooks}

In conclusion, we have performed the MDI-QKD experiment by using the state-of-the-art double-scanning method, drastically reducing the finite size effect. With an repetition rate of 50 MHz QKD system and run it for 5 minutes, we can obtain the secret key rates of 43.54 bps and 0.059 bps at 120 km and 150 km, respectively, which is impossible for all former methods. Expectably, if the repetition rate of the QKD system is increased to GHz level \cite{GHZ}, the time consumption will be further compressed to few seconds, which is very promising for one-time pad implementation of quantum communications. Therefore, our present work represent a further step towards practical implementation of MDI-QKD.
\\
\\
\noindent
\textbf{Funding.} This project supported by the National Key Research and Development Program of China (Grants Nos. 2018YFA0306400, 2017YFA0304100), the National Natural Science Foundation of China (Grants Nos.12074194, 11774180, U19A2075), the Leading-edge technology Program of Jiangsu Natural Science Foundation (Grants No. BK20192001), and the NUPTSF (Grants No. NY220122).
\\
\\
\textbf{Disclosures.} The authors declare no conflicts of interest.
\\
\\
$\dagger$ These two authors contribute equally to this work.

\end{document}